\documentclass{article}

\begin{document}

\title{Schr\"odinger equations for the square root density of an
eigenmixture and the square root of an eigendensity spin matrix}

\author{B. G. Giraud \\
bertrand.giraud@cea.fr, Institut de Physique Th\'eorique, \\
and\\
P. Moussa \\
pierre.moussa@cea.fr, Institut de Physique Th\'eorique, \\
DSM, CE Saclay, F-91191 Gif/Yvette, France}

\date{\today} 
\maketitle

\begin{abstract}

We generalize a ``one eigenstate'' theorem of Levy, Perdew and Sahni (LPS)
\cite{LevPerSah} to the case of densities coming from eigenmixture density 
operators. The generalization is of a special interest for the radial density 
functional theory (RDFT) for nuclei \cite{GirRDF}, a consequence of the 
rotational invariance of the nuclear Hamiltonian; when nuclear ground states
(GSs) have a finite spin, the RDFT uses eigenmixture density operators to 
simplify predictions of GS energies into one-dimensional, radial calculations.
We also study Schr\"odinger equations governing spin eigendensity matrices.

\end{abstract}

\medskip
The theorem of Levy, Perdew and Sahni \cite{LevPerSah} may be described as 
follows: 

\noindent
i) let $H_A$ be a Hamiltonian for $A$ identical particles, with individual 
mass $m,$
\begin{equation} 
H_A=\sum_{i=1}^A [ -\hbar^2 \Delta_{\vec r_i}/(2 m) + u(\vec r_i) ] + 
\sum_{i>j=1}^A v(\vec r_i,\vec r_j),
\end{equation}
ii) consider a GS eigenfunction of $H_A,$ 
$\psi(\vec r_1, \sigma_1, \vec r_2, \sigma_2, ...,\vec r_A, \sigma_A),$ 
where $\sigma_i$ denotes the spin state of the particle with space 
coordinates $\vec r_i,$ 

\noindent
iii) use a trace of $| \psi \rangle \langle \psi |$ upon all space 
coordinates but the last one, and upon all spins, to define the density,
\begin{equation}
\rho(\vec r) = A \sum_{\sigma_1...\sigma_A} \int d\vec r_1
d\vec r_2 ... d \vec r_{A-1} |\psi(\vec r_1, \sigma_1, \vec r_2, \sigma_2, 
..., \vec r_{A-1}, \sigma_{A-1}, \vec r, \sigma_A)|^2,
\end{equation}

\noindent
iv) then there exists a local potential $v_{eff}(\vec r)$ so that,
\begin{equation}
[-\hbar^2 \Delta_{\vec r}/(2 m)+v_{eff}(\vec r)]\, \sqrt{\rho(\vec r)}=
(E_A-E_{A-1})\, \sqrt{\rho(\vec r)}\, ,
\end{equation}
where the eigenvalue is the difference of the GS energy $E_A$ of the 
$A$-particle system and that, $E_{A-1},$ of the $(A-1)$-particle one.

Can this theorem be generalized for densities derived from eigenoperators,
${\cal D} \propto \sum_{n=1}^{\cal N} w_n | \psi_n \rangle \langle \psi_n |,$ 
corresponding to cases where $H$ has several $({\cal N}>1)$ degenerate 
GSs $\psi_n$? The degeneracy situation is of a wide interest in nuclear 
physics for doubly odd nuclei, the GSs of which often have a finite 
spin, and, if only because of Kramer's degeneracy, for odd nuclei. In 
particular, because of the rotational invariance of the nuclear Hamiltonian,
the density operator of interest for the RDFT \cite{GirRDF} reads,
${\cal D}=\sum_M | \psi_{JM} \rangle \langle \psi_{JM} |/(2J+1),$ where $J$ 
and $M$ are the usual angular momentum numbers of a degenerate magnetic 
multiplet of GSs $\psi_{JM}.$ Actually, more generally, it will easily 
be seen that the argument which follows holds for a degenerate multiplet of 
excited states as well.

This paper proves the generalization, by closely following the argument used
for one eigenstate only \cite{LevPerSah}. Furthermore, there is no need to
assume identical particles. No symmetry or antisymmetry assumption for 
eigenfunctions is needed. Let $\vec p_i$ and $\vec \sigma_i$ be the 
momentum and spin operators for the $i$th particle, at position $\vec r_i.$
Single out the $A$-th particle, with its degrees of freedom labelled 
$\vec r$ and $\vec \sigma$ rather than $\vec r_A$ and $\vec \sigma_A.$ 
For a theorem of maximal generality, with distinct masses, one-body and 
two-body potentials, our Hamiltonian may become, 
${\cal H}_A={\cal H}_{A-1}+{\cal V}_A+h_A,$ with
\begin{eqnarray} 
 {\cal H}_{A-1}&=&\sum_{i=1}^{A-1} [ -\hbar^2 \Delta_{\vec r_i}/(2 m_i) + 
u_i(\vec r_i, \vec p_i, \vec \sigma_i) ] + \sum_{i>j=1}^{A-1} 
v_{ij}(\vec r_i, \vec p_i, \vec \sigma_i,\vec r_j, \vec p_j, \vec \sigma_j)
\, , \nonumber \\
 {\cal V}_A&=&\sum_{j=1}^{A-1} 
v_{Aj}(\vec r,\vec r_j,\vec p_j, \vec \sigma_j)\, ,\ \ \ \ 
h_A= -\hbar^2 \Delta_{\vec r}/(2 m_A) + u_A(\vec r)\, .
\label{split} 
\end{eqnarray}
The potentials acting upon the first $(A-1)$ particles may be non local and 
spin dependent, but, for a technical reason which will soon become 
obvious, those potentials acting upon the $A$-th particle in ${\cal V}_A$ 
and $h_A$ must be strictly local and independent of the $A$-th spin. For 
notational simplicity, we choose units so that $\hbar^2/(2 m_A)=1$ from now on.

As in the one eigenstate case \cite{LevPerSah} we select situations where
there exists a representation in which, simultaneously, the Hermitian 
Hamiltonian ${\cal H}_A$ and all the eigenfunctions 
$\psi(\vec r_1, \sigma_1,..., \vec r_A, \sigma_A)$ under consideration 
are real. This reality condition does not seem to be restrictive, in view
of time reversal invariance.

Let $E_A$ be a degenerate eigenvalue of ${\cal H}_A.$ The degeneracy 
multiplicity being larger than $1,$ select ${\cal N} \ge 2$  of the 
corresponding eigenfunctions $\psi_n,$ orthonormalized. Their set may be 
either complete or incomplete in the eigensubspace. The density operators,
\begin{equation}
{\cal D} = \sum_{n=1}^{\cal N} | \psi_n \rangle\, w_n\,  \langle \psi_n |\, ,
\ \ \ \ \sum_{n=1}^{\cal N} w_n=1\, ,
\end{equation}
with otherwise arbitrary, positive weights $w_n,$ are normalized to unity, 
${\rm Tr}\, {\cal D}=1,$ in the $A$-body space. They are eigenoperators of 
${\cal H}_A,$ namely ${\cal H}_A\, {\cal D}=E_A\, {\cal D}.$

The partial trace of a ${\cal D}$ upon the first $(A-1)$ coordinates and all 
$A$ spins,
\begin{equation}
\tau(\vec r) = \sum_{n=1}^{\cal N} w_n 
\sum_{\sigma_1...\sigma_{A-1}\sigma} \int d\vec r_1 ... d \vec r_{A-1} 
[\psi_n(\vec r_1, \sigma_1,..., \vec r_{A-1}, \sigma_{A-1}, \vec r, \sigma)]^2,
\label{profile}
\end{equation}
defines a ``density'' $\tau$, normalized so that 
$\int d \vec r\ \tau(\vec r)=1.$ Let now $\phi_{n\, \vec r\, \sigma}$ be, in 
the space of the first $(A-1)$ particles, an auxiliary wave function defined 
by,
\begin{equation}
\phi_{n\, \vec r\, \sigma}(\vec r_1,\sigma_1,...,\vec r_{A-1},\sigma_{A-1})=
\psi_n(\vec r_1,\sigma_1,...,\vec r_{A-1},\sigma_{A-1},\vec r,\sigma)/
\sqrt{\tau(\vec r)}\, .
\end{equation}
Note that this auxiliary wave function depends on the choice of the 
weights $w_n.$ 

Now the density operator in the space of the first $(A-1)$ particles,
${\cal D}^{\, \prime}_{\vec r}=\sum_{n\sigma} 
|\phi_{n\, \vec r\, \sigma} \rangle\, w_n\, 
\langle \phi_{n\, \vec r\, \sigma} |,$ 
is normalized, ${\rm Tr}^{\, \prime}\, {\cal D}^{\, \prime}_{\vec r}=1,$ 
where the symbol ${\rm Tr}^{\, \prime}$ means integration upon the first 
$(A-1)$ coordinates and sum upon the first $(A-1)$ spins. Since this 
normalization of ${\cal D}^{\, \prime}_{\vec r}$ in the $(A-1)$-particle 
space does not depend on $\vec r,$ two trivial consequences read,
$\nabla_{\vec r}\, {\rm Tr}^{\, \prime}\, {\cal D}^{\, \prime}_{\vec r}=0$
and 
$\Delta_{\vec r}\, {\rm Tr}^{\, \prime}\, {\cal D}^{\, \prime}_{\vec r}=0.$ 
More explicitly, this gives,
\begin{eqnarray}
& \sum_{n\, \sigma\, \sigma_1...\sigma_{A-1}} \int d \vec r_1...d \vec r_{A-1} 
& w_n\ \, 
\phi_{n\, \vec r\, \sigma}(\vec r_1,\sigma_1,...,\vec r_{A-1},\sigma_{A-1}) 
\times \nonumber \\
&& \nabla_{\vec r}\
\phi_{n\, \vec r\, \sigma}(\vec r_1,\sigma_1,...,\vec r_{A-1},\sigma_{A-1})
= 0\, ,
\label{grad0}
\end{eqnarray}
and
\begin{eqnarray}
\sum_{n\, \sigma\, \sigma_1...\sigma_{A-1}} \int d \vec r_1...d \vec r_{A-1}
\  \, w_n\ \,  
\{\, [\nabla_{\vec r}\,
\phi_{n\, \vec r\, \sigma}(\vec r_1,\sigma_1,...,\vec r_{A-1},\sigma_{A-1})]^2 
&+& \nonumber \\
\phi_{n\, \vec r\, \sigma}(\vec r_1,\sigma_1,...,\vec r_{A-1},\sigma_{A-1}) 
\, \Delta_{\vec r}\,
\phi_{n\, \vec r\, \sigma}(\vec r_1,\sigma_1,...,\vec r_{A-1},\sigma_{A-1})\, 
\} &=& 0\, .
\label{lapl0}
\end{eqnarray}

Then one can rewrite the eigenstate property, $({\cal H}_A-E_A)\, \psi_n=0,$
into, 
\begin{equation}
({\cal H}_{A-1}+{\cal V}_A+h_A-E_A)\ \sqrt{\tau}\ \phi_{n\, \vec r\, \sigma}
=0 \, .
\label{eigen}
\end{equation}
This also reads,
\begin{eqnarray}
& \sqrt{\tau}\ 
({\cal H}_{A-1}+{\cal V}_A+u_A-E_A)\  \phi_{n\, \vec r\, \sigma} - 
(\Delta_{\vec r}\, \sqrt{\tau})\ \phi_{n\, \vec r\, \sigma} 
= & \nonumber \\
& 2\, (\nabla_{\vec r}\, \sqrt{\tau}) \cdot 
(\nabla_{\vec r}\, \phi_{n\, \vec r\, \sigma}) + 
\sqrt{\tau}\
(\Delta_{\vec r}\, \phi_{n\, \vec r\, \sigma})\, . &
\label{eigenagain}
\end{eqnarray}
The right-hand side (rhs) of Eq. (\ref{eigenagain}) occurs because the 
Laplacian, $\Delta_{\vec r},$ present in $h_A,$ acts also upon the parameter, 
$\vec r,$ of $\phi_{n\, \vec r\, \sigma}.$ This is where the local, 
multiplicative nature of $u_A$ and $v_{Aj}$ in the last particle space 
is used and avoids the occurrence of further terms, that would induce a 
somewhat unwieldy theory.

Define, for any integrand $\Psi_{n\vec r \sigma},$  the following
expectation value in the first $(A-1)$-particle space,
\begin{equation}
\langle \langle\Psi_{n\, \vec r\, \sigma} \rangle \rangle =
\sum_{\sigma_1...\sigma_{A-1}} \int d \vec r_1...d \vec r_{A-1}\,  
\Psi_{n\, \vec r\, \sigma}(r_1,\sigma_1,...,r_{A-1},\sigma_{A-1})\, .
\label{varnorm}
\end{equation}
Multiply Eq. (\ref{eigenagain}) by $\phi_{n\, \vec r\, \sigma}$ and 
integrate out the first $(A-1)$ coordinates and spins, to obtain,
\begin{eqnarray}
& \langle \langle\phi_{n\, \vec r\, \sigma}^2 \rangle \rangle\ 
[\, E^{exc}_{n \sigma}(\vec r)+E_{A-1}+U_{n \sigma}(\vec r)+ 
u_A(\vec r)-E_A-\Delta_{\vec r}\, ]\, \sqrt{\tau(\vec r)} & \nonumber \\
& =  2\, \langle \langle\ \phi_{n\, \vec r\, \sigma}\, 
(\nabla_{\vec r}\, \phi_{n\, \vec r\, \sigma})\ \rangle \rangle \cdot 
[\nabla_{\vec r}\, \sqrt{\tau(\vec r)}] +
\langle \langle\ \phi_{n\, \vec r\, \sigma}\, 
(\Delta_{\vec r}\, \phi_{n\, \vec r\, \sigma})\ \rangle \rangle\, 
\sqrt{\tau(\vec r)}\, , &
\label{traceprimeterm}
\end{eqnarray}
where $E^{exc}_{n \sigma}(\vec r)$ is defined from,
\begin{eqnarray}
&\langle \langle\phi_{n\, \vec r\, \sigma}^2 \rangle \rangle\
[E^{exc}_{n \sigma}(\vec r)+E_{A-1}] =
\sum_{\sigma_1...\sigma_{A-1}} \int d \vec r_1...d \vec r_{A-1}& \nonumber \\
&\phi_{n\, \vec r\, \sigma}(r_1,\sigma_1,...,r_{A-1},\sigma_{A-1})\,
\left[{\cal H}_{A-1}\, 
\phi_{n\, \vec r\, \sigma}\right](r_1,\sigma_1,...,r_{A-1},\sigma_{A-1})\, ,&
\label{excit}
\end{eqnarray} 
and $U_{n \sigma}(\vec r)$ results from,
\begin{eqnarray}
& \langle \langle\phi_{n\, \vec r\, \sigma}^2 \rangle \rangle\
U_{n \sigma}(\vec r) = \sum_{j=1}^{A-1} \sum_{\sigma_1...\sigma_{A-1}} 
\int d \vec r_1...d \vec r_{A-1}& \nonumber \\
&\phi_{n\, \vec r\, \sigma}(r_1,\sigma_1,...,r_{A-1},\sigma_{A-1}) 
v_{Aj}(\vec r,\vec r_j,\vec p_j,\vec \sigma_j) 
\phi_{n\, \vec r\, \sigma}(r_1,\sigma_1,...,r_{A-1},\sigma_{A-1})
 .&
\label{Hartree}
\end{eqnarray}

The square norm of $\phi_{n\, \vec r\, \sigma}$ in the $(A-1)$-particle 
space results from Eq. (\ref{varnorm}) with $\Psi=\phi^2.$ In 
Eq. (\ref{excit}) the expectation value of ${\cal H}_{A-1}$ is explicited as 
the sum of the GS energy $E_{A-1}$ of ${\cal H}_{A-1}$ and a positive, 
excitation energy $E^{exc}_{n \sigma}(\vec r).$ From Eq. (\ref{Hartree}), the 
Hartree nature of the potential $U_{n \sigma}(\vec r)$ is transparent.

Keeping in mind that, $\forall\, \vec r,$ the density operator 
${\cal D}^{\, \prime}$ is normalized to unity, namely, that 
$\sum_{n \sigma} w_n 
\langle \langle\phi_{n\, \vec r\, \sigma}^2 \rangle \rangle=1,$ multiply 
Eq. (\ref{traceprimeterm}) by $w_n$ and perform the sum upon $n$ and 
$\sigma.$ This gives,
\begin{eqnarray}
&[\, U^{exc}(\vec r)+E_{A-1}+U^{Hrt}(\vec r)+u_A(\vec r)-E_A -
\Delta_{\vec r}\, ]\ \sqrt{\tau} =  &
\nonumber \\
& \sum_{n \sigma} w_n\, 
\left[\, 2\, \langle \langle\, \phi_{n\, \vec r\, \sigma}\, 
(\nabla_{\vec r}\, \phi_{n\, \vec r\, \sigma})\, \rangle \rangle \cdot 
(\nabla_{\vec r}\, \sqrt{\tau}) + 
\langle \langle\, \phi_{n\, \vec r\, \sigma}\, 
(\Delta_{\vec r}\, \phi_{n\, \vec r\, \sigma})\, \rangle \rangle\,
 \sqrt{\tau}\, \right]\, , &
\label{traceprime}
\end{eqnarray}
where the ``mixture excitation potential'',
\begin{equation}
U^{exc}(\vec r)=\sum_{n \sigma}\, w_n\ 
\langle \langle \phi_{n\, \vec r\, \sigma}^2 \rangle \rangle\ 
E_{n \sigma}^{exc}(\vec r)\, ,
\end{equation}
is local and positive and the ``mixture Hartree-like potential'',
\begin{equation}
U^{Hrt}(\vec r) = \sum_{n \sigma}\, w_n\
\langle \langle \phi_{n\, \vec r\, \sigma}^2 \rangle \rangle\ 
 U_{n \sigma}(\vec r)\, ,
\end{equation}
is also local. Because of the frequent dominance of attractive terms in
$v_{Aj},$ it may show more negative than positive signs. Then notice that, 
because of Eqs. (\ref{grad0}), and Eq. (\ref{varnorm}) with 
$\Psi=\phi\, \nabla \phi,$ the sum in the rhs of Eq. (\ref{traceprime}),
$\sum_{n\sigma}\, w_n\ \langle \langle\, \phi_{n\, \vec r\, \sigma}\, 
(\nabla_{\vec r}\, \phi_{n\, \vec r\, \sigma})\, \rangle \rangle,$ vanishes. 
Note also, from Eqs. (\ref{lapl0}), and Eq. (\ref{varnorm}) with 
$\Psi=\phi\, \Delta \phi,$ that, again for the rhs of Eq. (\ref{traceprime}), 
the following equality holds,
\begin{equation}
- \sum_{n \sigma}\, w_n\ \langle \langle\, \phi_{n\, \vec r\, \sigma}\, 
(\Delta_{\vec r}\, \phi_{n\, \vec r\, \sigma})\, \rangle \rangle = 
\sum_{n\sigma}\, w_n\ \langle \langle\, 
(\nabla_{\vec r}\, \phi_{n\, \vec r\, \sigma}) \cdot 
(\nabla_{\vec r}\, \phi_{n\, \vec r\, \sigma})\, \rangle \rangle\, ,
\label{kin}
\end{equation}
where, again, the symbol $\langle \langle\ \rangle \rangle$ denotes the 
trace ${\rm T}r^{\, \prime},$ an integration upon the first $(A-1)$ 
coordinates together with summation upon their spins. The rhs of this 
equation, Eq. (\ref{kin}), defines a positive, local potential,
\begin{equation} 
U^{kin}(\vec r)=\sum_{n\sigma}\, w_n\ \langle \langle\, 
(\nabla_{\vec r}\, \phi_{n\, \vec r\, \sigma}) \cdot 
(\nabla_{\vec r}\, \phi_{n\, \vec r\, \sigma})\, \rangle \rangle\, .
\end{equation}

Finally, according to Eq. (\ref{traceprime}), the sum of local potentials, 
$U^{eff}=U^{exc}+U^{Hrt}+U^{kin}+u_A,$ drives a Schr\"odinger equation for 
$\sqrt \tau,$
\begin{equation}
[-\Delta_{\vec r}+U^{eff}(\vec r)]\, \sqrt{\tau(\vec r)} =
(E_A-E_{A-1})\, \sqrt{\tau(\vec r)}\, .
\label{genescal}
\end{equation}
This is the expected generalization of the LPS theorem. Note that, if the
$(A-1)$ particles are not identical, then $E_{A-1},$ the GS energy of 
${\cal H}_{A-1}$, means here the mathematical, absolute lower bound of the 
operator in all subspaces of arbitrary symmetry or lack of symmetry. In 
practice, however, most cases imply symmetries in the $(A-1)$-space, and 
$E_{A-1}$ means the ground state energy under such symmetries.

For nuclear physics, this generalization can be used in two ways:

\noindent
i) The first one consists in considering hypernuclei or mesonic nuclei, where 
the $A$-th particle is indeed distinct. Theoretical calculations with local 
interactions for the distinct particle may be attempted while non local 
and/or spin dependent interactions for the $A-1$ nucleons are useful, if not
mandatory. Then, obviously, the density $\tau$ refers to the hyperon or the 
meson and, given the neutron and proton respective numbers $N$ and $Z,$ wave 
functions $\psi_n$ and $\phi_{n\, \vec r,\, \sigma}$ belong to both $N$- and 
$Z$-antisymmetric subspaces. The energy $E_{A-1}$ is the GS energy of nucleus
$\{N,Z\},$ a fermionic GS energy, rather than the absolute lower bound of the
mathematical operator ${\cal H}_{A-1}$ in all subspaces.

\noindent 
ii) The second one consists in setting all $A$ particles to be nucleons, at 
the cost of restricting theoretical models to local interactions. Such models 
are not without interest indeed, although interactions which are spin 
dependent are certainly more realistic. The antisymmetric properties of the 
functions $\psi_n$  are requested in both N- and Z-spaces. If the singled 
out, $A$-th coordinate is a neutron one, the density $\tau$ defined by 
Eq. (\ref{profile}) is the usual neutron density, divided by $N;$ the
functions $\phi_{n\, \vec r\, \sigma}$ are antisymmetric in the 
$(N-1)$-neutron space and the $Z$-proton space; the energy $E_{A-1}$ now 
means the fermionic GS of nucleus $\{N-1,Z\},$ not that absolute, 
mathematical lower bound of operator ${\cal H}_{N-1,Z}.$  Conversely, if 
the $A$-th coordinate is a proton one, then, {\it mutatis mutandis}, $\tau$ 
is the usual proton density, divided by $Z,$ and $E_{A-1}$ is the GS energy 
of nucleus $\{N,Z-1\}.$

In both cases, the Hamiltonians to be used are scalars under the rotation 
group, and, therefore \cite{GirRDF}, the density operators ${\cal D}$ 
considered by the RDFT are also scalars. Hence, all calculations defining 
$U^{eff}$ reduce to calculations with a radial variable $r$ only.

We shall now extend our previous results to the case where we allow spin
dependence for all interactions, a most useful feature if all $A$ particles
are nucleons. Polarized eigenmixtures are also interesting and need also be
considered. Hence, a generalisation of our approach, which uses the 
``spin-density matrix'' (SDM) formalism \cite{GunLun} \cite{Goer1}, is in 
order. The Hamiltonian may become,
\begin{equation} 
\sum_{i=1}^A [ -\hbar^2 \Delta_{\vec r_i}/(2 m_i) + 
u_i(\vec r_i,\vec \sigma_i) ] + 
\sum_{i>j=1}^A v_{ij}(\vec r_i,\vec \sigma_i,\vec r_j,\vec \sigma_j)\, .
\end{equation}
It allows subtle differences between neutrons and protons, besides the 
Coulomb interactions between protons. More explicitly, there can be two 
distinct one-body potentials, $u_n,u_p,$ namely one for neutrons and one for 
protons, but within the neutron space the function $u_n(\vec r_i,\sigma_i)$ 
obviously will not read $u_{ni}(\vec r_i,\sigma_i).$ Similarly in the proton 
space, the Hamiltonian contains terms $u_p(\vec r_i,\sigma_i)$ rather than 
$u_{pi}({\vec r_i,\sigma_i}.$
The same subtlety allows terms 
$v_{pp}(\vec r_i,\sigma_i,\vec r_j,\sigma_j),$
$v_{pn}(\vec r_i,\sigma_i,\vec r_j,\sigma_j),$
$v_{np}(\vec r_i,\sigma_i,\vec r_j,\sigma_j)$ and
$v_{pp}(\vec r_i,\sigma_i,\vec r_j,\sigma_j),$
rather than 
$v_{ppij}(\vec r_i,\sigma_i,\vec r_j,\sigma_j),$ ... etc. (Of course, 
$v_{pn}=v_{np}.$)
But non localities of potentials and interactions, in the sense of explicit 
dependences upon momenta $p_i,$ remain absent. 

Then the $A$-th particle is again singled out, with degrees of freedom again 
labelled $\vec r$ and $\vec \sigma,$ and the Hamiltonian is split as a sum, 
${\cal K}_{A-1}+{\cal W}_A+k_A,$ somewhat similar to the split described by 
Eqs. (\ref{split}). For simplicity, we shall use short notations, 
${\cal K},$ ${\cal W}$ and $k,$ rather than ${\cal K}_{A-1},$ ${\cal W}_A$ 
and $k_A.$ With two spin states, $\sigma =\pm 1/2,$ for the $A$-th nucleon, 
we represent eigenstates $\psi_n$ as column vectors, 
$\overline \psi_n=\left[\matrix{\psi_{n+} \cr \psi_{n-}}\right],$ and 
operators as matrices such as, 
$\overline {\cal W}=\left[\matrix{
{\cal W}_{++} & {\cal W}_{+-} \cr {\cal W}_{-+} & {\cal W}_{--}}\right],$  
$\overline k= \left[\matrix{k_{++} & k_{+-} \cr k_{-+} & k_{--}}\right]$ and
$\bar u= \left[\matrix{u_{A++} & u_{A+-} \cr u_{A-+} & u_{A--}}\right].$
The matrix, 
$\overline {\cal K}=\left[\matrix{{\cal K} & 0 \cr 0 & {\cal K}}\right],$ 
is a scalar in spin space, since ${\cal K}$ does not act upon the $A$-th 
particle. 

The spin density matrix, $\overline \rho_n,$ results from an integration and 
spin sum over the $(A-1)$-space of the matrix, 
$\overline \psi_n \, \times\, \overline \psi_n^T,$ where the superscript $^T$ 
denotes transposition,
\begin{equation}
\overline \rho_n(\vec r)=
\langle \langle\ \left[\matrix{\psi_{n+}  \cr \psi_{n-} }\right]\, \times\,
                \left[\matrix{\psi_{n+}  &  \psi_{n-}}\right]\ \rangle \rangle 
=
\langle \langle 
\left[\matrix{(\psi_{n+})^2  & \psi_{n+}  \psi_{n-} \cr 
              \psi_{n-}  \psi_{n+} & (\psi_{n-})^2 }\right] 
\rangle \rangle\, .
\end{equation}
It depends on the last coordinate, $\vec r,$ and its matrix elements are 
labelled by two values, $\{\sigma,\sigma'\},$ of the last spin.
For an eigenmixture one defines, obviously, 
$\bar \theta(\vec r)=\sum_n w_n\, \overline \rho_n(\vec r),$ and the 
trace in the last spin space, $[\theta_{++}(\vec r)+\theta_{--}(\vec r)],$ is 
that density, $\tau(\vec r),$ defined by Eq. (\ref{profile}).

The SDM, $\bar \theta,$ is symmetric and positive semidefinite, 
$\forall\, \vec r.$ Except for marginal situations, it is
also invertible, in which case there exists a unique inverse square root, also
symmetric and positive. Define, therefore, a column vector 
$\overline \phi_{n\, \vec r}$ of states in the $(A-1)$-space according to,
\begin{equation}
\overline \phi_{n\, \vec r}= 
\bar \theta^{\, -\frac{1}{2}}(\vec r)\ \overline \psi_n\, .
\end{equation}
Then the following property, 
\begin{equation}
\sum_n w_n\, \langle \langle\, \overline \phi_{n\, \vec r} \times 
\overline \phi_{n\, \vec r}^T\, 
\rangle \rangle = \bar \theta^{\, -\frac{1}{2}}(\vec r)\ \left( 
\sum_n\, w_n\, \langle \langle\ \overline \psi_n \times \overline \psi_n^T\ 
\rangle \rangle \right) \bar \theta^{\, -\frac{1}{2}}(\vec r) = \bar {\bf 1}
\, ,
\label{identmatr}
\end{equation}
holds $\forall\, \vec r.$ Here $\bar {\bf 1}$ denotes the identity matrix. 
Hence, the following gradient and Laplacian properties also hold,
\begin{eqnarray}
\sum_n w_n\,\langle \langle\ 
\left( \nabla_{\vec r} \left[\matrix{\phi_{n\, \vec r\, +}  \cr 
\phi_{n\, \vec r\, -} }  \right] \right) \times  
\left[ \matrix{ \phi_{n\, \vec r\, +} & \phi_{n\, \vec r\, -} } \right] 
\ \rangle \rangle &+& \nonumber \\
\sum_n w_n\,\langle \langle\ 
 \left[\matrix{\phi_{n\, \vec r\, +}  \cr 
\phi_{n\, \vec r\, -} }  \right] \times  
\left(\ \nabla_{\vec r}\ 
\left[ \matrix{ \phi_{n\, \vec r\, +} & \phi_{n\, \vec r\, -} } \right]\  
\right)\ \rangle \rangle
&=&0\, ,\ \ \ \ \forall\, \vec r\, ,
\label{nulgrad}
\end{eqnarray}

and 

\begin{eqnarray}
& \sum_n w_n\, \langle \langle\ 
\left( \Delta_{\vec r} \left[\matrix{\phi_{n\, \vec r\, +}  \cr 
\phi_{n\, \vec r\, -} }  \right] \right) \times  
\left[ \matrix{ \phi_{n\, \vec r\, +} & \phi_{n\, \vec r\, -} } \right] 
\ \rangle \rangle\ + & 
\nonumber \\
& 2\, \sum_n w_n\, \langle \langle\ 
\left( \nabla_{\vec r} \left[\matrix{\phi_{n\, \vec r\, +}  \cr 
\phi_{n\, \vec r\, -} }  \right] \right) \cdot  
\left( \nabla_{\vec r} \left[ \matrix{ \phi_{n\, \vec r\, +} & 
\phi_{n\, \vec r\, -} } \right] \right)
\ \rangle \rangle\ + &
\nonumber \\
& \sum_n w_n\, \langle \langle\ 
\left[\matrix{\phi_{n\, \vec r\, +}  \cr \phi_{n\, \vec r\, -} } \right]
\times \left(\, \Delta_{\vec r}
\left[ \matrix{ \phi_{n\, \vec r\, +} & \phi_{n\, \vec r\, -} } \right]\ 
\right)\ \rangle \rangle = 0\, ,\ \ \ \ \forall\, \vec r&
\, .
\label{nullapl}
\end{eqnarray}

The eigenvector property, 
$\left( \overline {\cal K}+\overline {\cal W}+\overline k - 
E_A\, \bar {\bf 1}\right)\, \overline \psi_n =0,$ also
reads,
\begin{eqnarray}
& \left( \overline {\cal K}+\overline {\cal W}+\bar u - E_A\, \bar {\bf 1} 
\right)\, \bar \theta^{\frac{1}{2}}(\vec r)\, \overline \phi_{n \, \vec r} - 
\left[ \Delta_{\vec r}\ \bar \theta^{\frac{1}{2}}(\vec r) \right]\, 
\overline \phi_{n \vec r} = & \nonumber \\
& 2 \left[ \nabla_{\vec r}\ \bar \theta^{\frac{1}{2}}(\vec r) \right] 
\cdot \left( \nabla_{\vec r}\ \overline \phi_{n\, \vec r} \right) + 
\bar \theta^{\frac{1}{2}}(\vec r)\ 
\left( \Delta_{\vec r}\ \overline \phi_{n\, \vec r} \right)\, .&
\label{matrixeigen}
\end{eqnarray}
Right-multiply Eq. (\ref{matrixeigen}) by the row vector, 
$\overline \phi_{n\, \vec r}^T\, ,$ integrate and sum over the $(A-1)$-space,
weigh the result by $w_n$ and sum upon $n.$ Because of Eq. (\ref{identmatr}), 
the weighted sum of averages over the $(A-1)$-space simplifies into,
\begin{eqnarray}
& \left[\overline {\cal U}^{exc}(\vec r)+(E_{A-1}-E_A)\, \bar {\bf 1}+
\overline {\cal U}^{Hrt}(\vec r)+\bar u(\vec r)-\Delta_{\vec r}\right]\, 
\bar \theta^{\frac{1}{2}}(\vec r)= \sum_n w_n \times & \nonumber \\
& \left\{\, 2 \left[ \nabla_{\vec r}\ \bar \theta^{\frac{1}{2}}(\vec r)
 \right] \cdot \, \langle \langle\,
\left( \nabla_{\vec r}\ \overline \phi_{n\, \vec r} \right) \times 
\overline \phi_{n\, \vec r}^T\, \rangle \rangle + 
\bar \theta^{\frac{1}{2}}(\vec r)\, \langle \langle\,
\left( \Delta_{\vec r}\, \overline \phi_{n\, \vec r} \right) \times 
\overline \phi_{n\, \vec r}^T\, \rangle \rangle\, \right\}\, ,&
\label{roughschroe}
\end{eqnarray}
with
\begin{equation}
\overline {\cal U}^{exc}(\vec r)=\sum_n w_n\ \langle \langle\  
\left( \overline {\cal K}\ \overline \phi_{n\, \vec r} \right) \times
\overline \phi_{n\, \vec r}^T\ \rangle \rangle - E_{A-1}\, \bar {\bf 1}\, ,
\end{equation}
and
\begin{equation}
\overline {\cal U}^{Hrt}(\vec r)=\sum_n w_n\ \langle \langle\  
\left( \overline {\cal W}\ \overline \phi_{n\, \vec r} \right) \times 
\overline \phi_{n\, \vec r}^T\ \rangle \rangle\, .
\end{equation}
The rhs of Eq. (\ref{roughschroe}) can be simplified, but less than that of
Eq. (\ref{traceprime}). Indeed Eq. (\ref{nulgrad}) does not imply that the
coefficient of $\nabla_{\vec r}\, \bar \theta^{\frac{1}{2}}(\vec r)$ vanishes. 
In fact this coefficient is,
\begin{equation}
\sum_n w_n 
\langle \langle \left( \nabla_{\vec r}\ \overline \phi_{n \vec r} \right) 
\times \overline \phi_{n \vec r}^T \rangle \rangle = \sum_n w_n 
\left[ \matrix{ 0 & \langle \langle (\nabla_{\vec r} 
\phi_{n \vec r +}) \phi_{n \vec r -} \rangle \rangle \cr 
\langle \langle (\nabla_{\vec r} \phi_{n \vec r -})\,
\phi_{n \vec r +} \rangle \rangle & 0} \right],
\label{antigrad}
\end{equation}
and Eq. (\ref{nulgrad}) shows that the matrix on the rhs is antisymmetric.
In turn, from Eq. (\ref{nullapl}), the  ``Laplacian induced 
coefficient'' in the rhs of Eq. (\ref{roughschroe}) may be listed as,
\begin{eqnarray}
& \sum_n w_n\, \langle \langle\, \left( \Delta_{\vec r}\,
\overline \phi_{n\, \vec r} \right) \times 
\overline \phi_{n\, \vec r}^T\, \rangle \rangle = & \nonumber \\
& - \sum_n w_n\, \langle \langle\,
\left( \nabla_{\vec r} \left[\matrix{\phi_{n\, \vec r\, +}  \cr 
\phi_{n\, \vec r\, -} }  \right] \right) \cdot  
\left( \nabla_{\vec r} \left[ \matrix{ \phi_{n\, \vec r\, +} & 
\phi_{n\, \vec r\, -} } \right] \right)\, \rangle \rangle\ + \frac{1}{2} 
\sum_n w_n\, \times & \nonumber \\
& \langle \langle\, \left[
\left( \Delta_{\vec r} \left[\matrix{\phi_{n \vec r +}  \cr 
\phi_{n \vec r -} }  \right] \right) \times  
\left[ \matrix{ \phi_{n \vec r +} & \phi_{n \vec r -} } \right] - 
\left[\matrix{\phi_{n \vec r +}  \cr \phi_{n \vec r -} } \right] \times
 \left( \Delta_{\vec r}
\left[ \matrix{ \phi_{n \vec r +} & \phi_{n \vec r -} } \right]
\right) \right]\, \rangle \rangle. &
\label{antilapl}
\end{eqnarray}
In the rhs of Eq. (\ref{antilapl}), the similarity of its first term with 
potential $U^{kin},$ Eq. (\ref{kin}), is transparent. Also transparent is
the antisymmetry of the second term. With the definitions,
\begin{equation}
\overline {\cal U}^{kin}(\vec r)=\sum_n w_n\, \langle \langle\,
\left( \nabla_{\vec r}\, \overline \phi_{n\, \vec r} \right) \cdot  
\left( \nabla_{\vec r}\, \overline \phi_{n\, \vec r}^T \right)\, 
\rangle \rangle\, ,
\end{equation}
\begin{equation}
\overline {\cal U}^{ant}(\vec r)=\frac{1}{2} \sum_n w_n\, \langle \langle\, 
\left[ \left( \Delta_{\vec r}\, \overline \phi_{n\, \vec r} \right) \times  
\overline \phi_{n\, \vec r}^T - \overline \phi_{n\, \vec r} \times
 \left( \Delta_{\vec r}\, \overline \phi_{n\, \vec r}^T \right) \right]\, 
\rangle \rangle\, ,
\end{equation}
and
\begin{equation}
\overline {\cal U}^{grd}(\vec r)=2 \sum_n w_n 
\left[ \matrix{ 0 & \langle \langle\, (\nabla_{\vec r}\, 
\phi_{n \vec r +})\, \phi_{n \vec r -}\, \rangle \rangle \cr 
\langle \langle\, (\nabla_{\vec r}\, \phi_{n \vec r -})\,
\phi_{n \vec r +}\, \rangle \rangle & 0} \right]\, ,
\end{equation}
the Schr\"odinger equation for the square root of the spin density matrix 
reads,
\begin{equation}
\left[\overline {\cal U}^{exc}+\overline {\cal U}^{Hrt}+\bar u
+\overline {\cal U}^{kin}-\overline {\cal U}^{ant}-\Delta\right]\, 
\bar \theta^{\frac{1}{2}}-\overline {\cal U}^{grd} \cdot \nabla 
\bar \theta^{\frac{1}{2}}=(E_A-E_{A-1})\, \bar \theta^{\frac{1}{2}}.
\label{finerschroe}
\end{equation}

This concludes our generalizations of the LPS theorem. On the one hand, see 
Eq. (\ref{genescal}), we obtained  for the square root density of an 
eigenmixture a Schr\"odinger equation, most similar to the LPS equation. 
On the other hand, at the cost of a slightly less simple result, we also
obtained, see Eq. (\ref{finerschroe}), an LPS-like equation that drives the 
square root of the spin density matrix. 

This opens a completely new zoology of local, effective potentials, of which 
very little is known, but the interest of which is obvious, since they drive 
a reasonably easily measurable observable, the square root of an eigenmixture 
density, which depends on one degree of freedom $\vec r$ only. A connection 
of such potentials with optical potentials, or rather their real parts, is 
likely, but yet remains an open problem.

\end{document}